\RequirePackage{amsthm}

\documentclass[sn-mathphys,Numbered]{sn-jnl}

\usepackage{tabularx,booktabs}
\newcolumntype{Y}{>{\centering\arraybackslash}X}
\usepackage{graphicx}%
\usepackage{pgfplots}
\usepackage[]{lineno}
\usepgfplotslibrary{colorbrewer}
\pgfplotsset{compat = 1.15, cycle list/Set1-8} 
\usepgfplotslibrary{external,colormaps,groupplots,statistics}

\usetikzlibrary{pgfplots.statistics, pgfplots.colorbrewer} 
\usepackage{pgfplotstable}
\usepackage{filecontents}
\usepackage{comment}

\usepackage{footnote}
\makesavenoteenv{algorithm}

\usepgfplotslibrary{statistics}
\usepackage{placeins}
\usepackage{multirow}%
\usepackage{amsmath,amssymb,amsfonts}%
\usepackage{amsthm}%
\usepackage{mathrsfs}%
\usepackage[title]{appendix}%
\usepackage{xcolor}%
\usepackage{textcomp}%
\usepackage{manyfoot}%
\usepackage{booktabs}%
\usepackage{algorithm}%
\usepackage{algpseudocode}%
\usepackage{float}
\usepackage{listings}%
\usepackage[acronym]{glossaries} 
\glsdisablehyper

\newacronym[shortplural={D$_{\text{dye}}$}, longplural={donor dye, ex. Alexa 488}]{ddye}{D$_{\text{dye}}$}{donor dye, ex. Alexa 488}

\newacronym{US}{US}{Ultrasound}
\newacronym{DL}{DL}{deep learning}
\newacronym{CT}{CT}{computer tomography}
\newacronym{MRI}{MRI}{magnetic resonance imaging}

\newacronym{STOA}{STOA}{state-of-the-art}
\newacronym{TRE}{TRE}{target registration error}
\newacronym{OR}{OR}{operating room}
\newacronym{FL}{FL}{x-ray fluoroscopy}
\newacronym{CAD}{CAD}{computer-aided design}
\newacronym{AR}{AR}{augmented reality}
\newacronym{VR}{VR}{virtual reality}
\newacronym{MLP}{MLP}{multi-layer perceptron}
\newacronym{SDF}{SDF}{signed distance function}
\newacronym{PCN}{PCN}{Point Completion Network}
\newacronym{MSN}{MSN}{Morphing and Sampling Network}
\newacronym{FinerPCN}{FinerPCN}{Finer Point Completion Network}
\newacronym{ViPC}{ViPC}{View-Guided Point Cloud Completion}
\newacronym{PointCNN}{PointCNN}{Point Convolutional Neural Network}
\newacronym[plural=CNNs,firstplural=convolutional neural networks (CNNs)]{CNN}{CNN}{convolutional neural network}
\newacronym[plural=IFNs,firstplural=Implicit Feature Networks]{IFN}{IFN}{Implicit Feature Network}
\newacronym{PPE}{PPE}{Point Positional Encoding}
\newacronym{GRNet}{GRNet}{Gridding Residual Network }
\newacronym{VE-PCN}{VE-PCN}{Voxel Edge Point Completion Network}
\newacronym{MRAC-Net}{MRAC-Net}{Multi-resolution Anisotropic Convolutional Network}
\newacronym[plural=VAEs, firstplural= Variational Autoencoders (VAEs)]{VAE}{VAE}{Variational Autoencoder}
\newacronym[plural=GANs, firstplural= Generative Adversarial Networks (GANs)]{GAN}{GAN}{generative adversarial network}
\newacronym{AE}{AE}{auto encoder}
\newacronym{PCT}{PCT}{Point Completion Transformer}
\newacronym{SA-Net}{SA-Net}{Shuffle Attention Network}
\newacronym{DSC}{DSC}{dice score}
\newacronym{HD}{HD}{Hausdorf Distance}
\newacronym{VerSe20}{VerSe20}{Large Scale Vertebrae Segmentation Challenge Dataset 2020}
\newacronym{PMNet}{PMNet}{Probabilistic Modeling Network}
\newacronym{RENet}{RENet}{Relational Enhacement Network }
\newacronym{CD}{CD}{Chamfer Distance}
\newacronym{EMD}{EMD}{Earth Mover's Distance}
\newacronym{PC}{PC}{point cloud}

\newacronym{F1}{F1}{F1-score}
\newacronym{VRCNet}{VRCNet}{Variational Relational Completion Network}
\newacronym{GT}{GT}{ground truth}
\newacronym{SOTA}{SOTA}{state-of-the-art}
\newacronym{KL}{KL}{Kullback-Leibler}
\newacronym{CL}{CL}{Chamfer Loss}
\newacronym{R-PSK}{R-PSK}{Residual Point Selective Kernel Module} 
\definecolor{green}{rgb}{0,0.85,0}

\newcommand{\revisionAdd}[1]{\textcolor{black}{#1}}
\newcommand{\revisionRemove}[1]{}

\algdef{SE}[SUBALG]{Indent}{EndIndent}{}{\algorithmicend\ }%
\algtext*{Indent}
\algtext*{EndIndent}

\theoremstyle{thmstyleone}%
\theoremstyle{thmstyletwo}%

\theoremstyle{thmstylethree}%

\begin{document}

\title[Article Title]{Shape Completion in the Dark: Completing
Vertebrae Morphology from 3D Ultrasound}


\author*[1]{\fnm{Miruna-Alexandra} \sur{Gafencu}}\email{miruna.gafencu@tum.de}
\author[1]{\fnm{Yordanka} \sur{Velikova}}
\author[1]{\fnm{Mahdi} \sur{Saleh}}
\author[2]{\fnm{Tamas} \sur{Ungi}}
\author[1]{\fnm{Nassir} \sur{Navab}}
\author[3,1]{\fnm{Thomas} \sur{Wendler}}
\author[1]{\fnm{Mohammad Farid} \sur{Azampour}}


\affil*[1]{\orgdiv{Computer-Aided Medical Procedure and Augmented Reality (CAMP), CIT}, \orgname{Technical University of Munich},\city{Garching bei Muenchen}, \country{Germany}}

\affil[2]{\orgdiv{School of Computing}, \orgname{Queen's University}, \orgaddress{\city{Kingston}, \state{Ontario}, \country{Canada}}}


\affil[3]{\orgdiv{Clinical Computational Medical Imaging Research, Department of Diagnostic and Interventional Radiology and Neuroradiology}, \orgname{University Hospital Augsburg}, \city{Augsburg}, \country{Germany}}



\abstract{
\textbf{Purpose:} 
\gls{US} imaging, while advantageous for its radiation-free nature, is challenging to interpret due to only partially visible organs and a lack of complete 3D information. While performing US-based diagnosis or investigation, medical professionals therefore create a mental map of the 3D anatomy. In this work, we aim to replicate this process and enhance the visual representation of anatomical structures.

 
\textbf{Methods:}
We introduce a point-cloud-based probabilistic \gls{DL} method to complete occluded anatomical structures through 3D shape completion and choose US-based spine examinations as our application. To enable training, we generate synthetic 3D representations of partially occluded spinal views by mimicking US physics and accounting for inherent artifacts.

 \textbf{Results:} 
The proposed model performs consistently on synthetic and patient data, with mean and median differences of 2.02 and 0.03 in \gls{CD}, respectively. Our ablation study demonstrates the importance of US physics-based data generation, reflected in the large mean and median difference of 11.8 \gls{CD} and 9.55 \gls{CD}, respectively. Additionally, we demonstrate that anatomic landmarks, such as the spinous process (with reconstruction \gls{CD} of 4.73) and the facet joints (mean distance to \gls{GT} of 4.96mm) are preserved in the 3D completion. 

\textbf{Conclusion:} 
Our work establishes the feasibility of 3D shape completion for lumbar vertebrae, ensuring the preservation of level-wise characteristics and successful generalization from synthetic to real data. The incorporation of US physics contributes to more accurate patient data completions. Notably, our method preserves essential anatomic landmarks and reconstructs crucial injections sites at their correct locations. The generated data and source code will be made publicly available~\footnote[1]{https://github.com/miruna20/Shape-Completion-in-the-Dark}.
}


\maketitle







\section{Introduction}

\gls{US} imaging provides a non-invasive, radiation-free, and low-cost way to observe internal structures and organs in real-time. While valuable, this modality has its own limitations such as reduced field of view, user dependence, and the presence of artifacts. 

Due to the underlying physical properties of US imaging, highly reflective structures such as bones introduce shadows occluding tissue below them.  In contrast, imaging techniques like CT and MRI provide comprehensive representations of anatomical structures without angle dependence and significantly fewer occlusion artifacts. Consequently, interpreting US images can be notably more challenging~\cite{azampour2023anatomy}.

When using US in a conventional fashion to extract anatomic information needed for diagnosis or intervention, medical professionals must rely on their expertise to mentally reconstruct the 3D shape of the organ or structure from partial US views. 
This not only adds to the time and effort of the diagnostic process but also presents a learning challenge for young professionals. Our objective is to assist in this process by \revisionAdd{enhancing the ultrasound view with the complete 3D shape and facilitating a rapid and more intuitive understanding of the anatomy}. We employ 3D shape completion techniques~\cite{guo2020deep} to deduce the complete contour of organs based on the partially visible anatomy in an US sweep, ensuring that salient structural details are preserved. In this manner, not only do we assist professionals, but we also translate this intricate cognitive task into a format 
machines can process. 


Various deep learning (DL) techniques have been proposed for 3D shape completion. These methods use a combination of local and global features with diverse representations. The early methods, like Point Completion Network (PCN)\cite{yuan2018pcn} and TopNet\cite{tchapmi2019topnet}, employ folding operations, offering a rough reconstruction of the shape modeled as a \gls{PC}. Later, DeepSDF\cite{park2019deepsdf} proposed to leverage continuous signed distance fields to learn about shape categories and improve quality. PoinTr\cite{yu2021pointr} then introduced a technique to predict only the missing region and concatenate the inputs and outputs of the model to produce the final completion. In one of the newer methods, \gls{VRCNet}\cite{pan2021variational}, a probabilistic approach is adopted. Here, a shape prior distribution is learned across various object classes. Following this learning, the shape completion is derived using Maximum a Posteriori (MAP) estimation, where the input partial \gls{PC} serves as the observed data.

This state-of-the-art shows that training DL methods for shape completion requires substantial datasets to achieve optimal results. 
In the computer vision realm, where all the cited work originates, CAD models of objects are often employed to produce realistic occlusions, thereby creating extensive training datasets. 
Extending this paradigm to the medical domain, our objective is to generate CAD-inspired representations of partial views in US by simulating physics-based occlusions. Generating synthetic data in this manner holds particular promise in medical areas where US is limited in clinical settings, where patient data is scarce or hard to get.

One such area is the examination and intervention on the spine, which is extensively explored in research but has yet to be fully established in clinical practice using US~\cite{gueziri2020state} despite its high potential to reduce radiation exposure to patients and medical personnel. The challenge with spine US scans lies in their limited visibility; only the posterior surface of the spine can be imaged. These scans are primarily affected by acoustic shadowing, preventing the US beam from reaching deeper vertebral structures below the spine surface. This limitation complicates the operator's comprehension of the entire spine anatomy.

Shape completion of the partially visible vertebra can help in overcoming this limitation. Inspired by \gls{VRCNet}, we propose to use a point-cloud-based probabilistic method that takes advantage of preexisting 3D imaging, such as \gls{CT}, which offers comprehensive 3D shape details, to understand shape priors. The proposed method learns fine 3D \gls{PC} geometries of vertebrae and predicts consistent and detailed \gls{PC}s for the occluded regions.

For the training of our model, we introduce a unique, fully automated pipeline for generating synthetic data. We generate physics-based synthetic data that mimics US characteristics, bridging and facilitating the application of \revisionAdd{diverse} shape completion techniques in medical contexts, \revisionAdd{otherwise unfeasible due to lack of access to paired \gls{US}/\gls{CT} data.} When integrated with the proposed 3D \gls{PC} reconstruction network, our pipeline enables the completion of vertebrae shapes from 3D US data. \revisionAdd{ Through shape completion we introduce a new perspective to tackle US data interpretability, and to the best of our knowledge propose a first work in the direction of 3D anatomic shape completion from ultrasound scans.}  In summary, the contributions are three-fold:


\begin{enumerate}
     \item We develop a synthetic data generation pipeline that produces realistic, US-consistent partial views of lumbar vertebrae. 
     \item We introduce a 3D shape completion pipeline for lumbar vertebrae.  
     \item We evaluate our method's shape completion capabilities on synthetic and CT-US patient data, and report standard computer vision metrics as well as anatomy-based ones. 
\end{enumerate}

\section{Materials and Methods}

\subsection{Synthetic Data Generation}~\label{section:synth_data_generation}
In a common computer vision pipeline, training data for shape completion is created by generating realistic occlusions of objects using CAD models, e.g., by ray-casting from different camera positions around the object. Much like this approach, our synthetic data generation pipeline utilizes high-resolution abdominal CT scans with vertebral masks to generate partial \gls{PC}s resembling vertebrae visibility in US. 



\begin{figure}[t]
    \centering
    \includegraphics[width=0.9\textwidth, keepaspectratio]{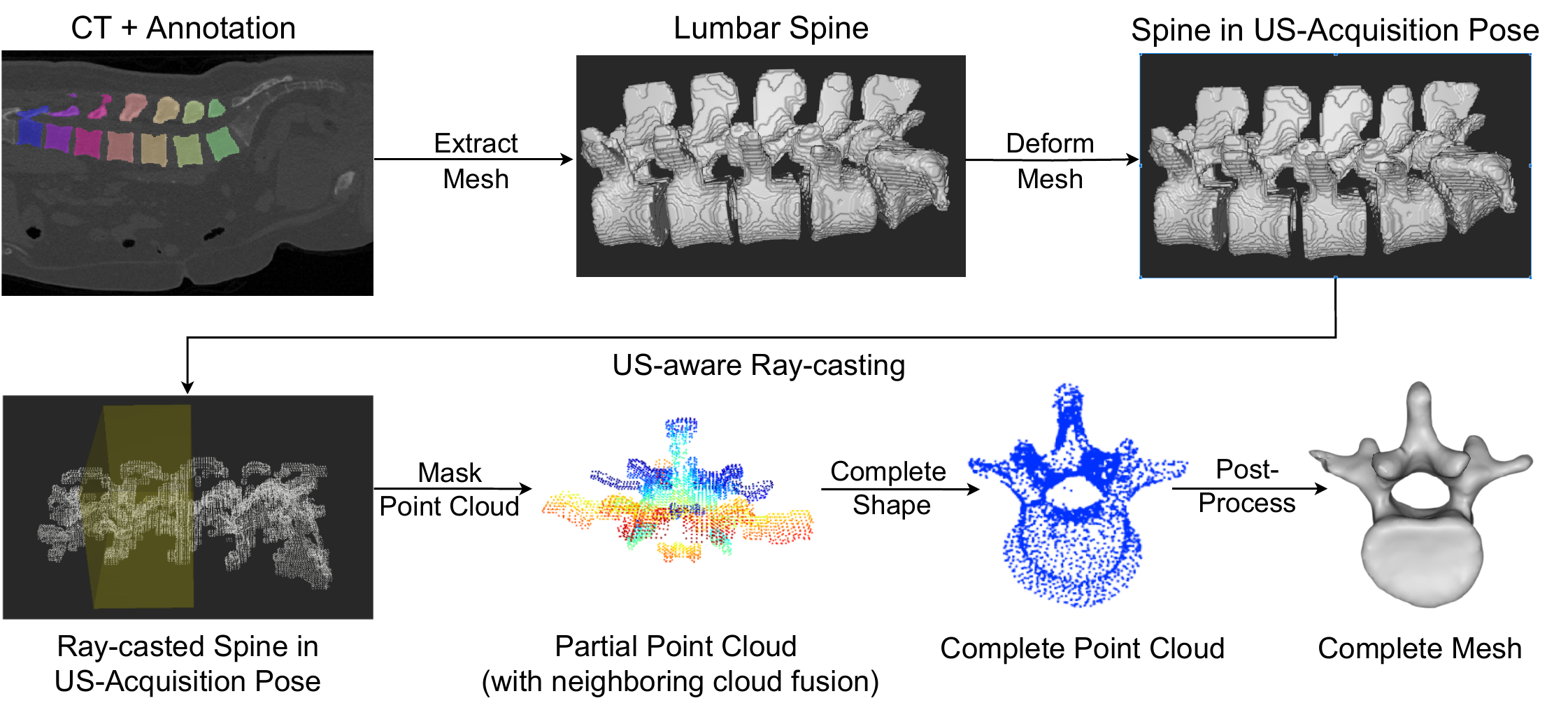}
    \caption{Overview of the training pipeline of our proposed method. First data generation is performed, followed by shape completion and post processing.}
    \label{fig:data_generation_pipeline}
\end{figure}

Three main milestones need to be achieved to generate a large amount of realistic, US-consistent partial views of the vertebrae only using an abdominal CT scan. First, we need to account for the multitude of possible patient positioning during the US acquisition. Second, we need to generate partial views of the spine that adhere to spine US acquisition techniques and their field of view and faithfully replicate the effects of US-characteristic artifacts such as acoustic shadowing or scattering. Lastly, our method must account for potential inaccuracies stemming from the error-prone, challenging task of vertebrae classification and annotation in US. Figure~\ref{fig:data_generation_pipeline} displays the complete data generation pipeline described in detail in the following. \revisionAdd{For an algorithmic overview please refer to the supplementary material.} 

\FloatBarrier

\subsubsection{Accounting for multiple spine curvatures }\label{subsection:Spine Deformation}
Patient positions during ultrasound screening vary depending on the target anatomy and spine region. While sitting is typical for visualizing the interlaminar space, the prone position aids lumbar facet joint access. To encompass this range of spine curvatures, we enhance the spine meshes from CT to produce varied realistic curvatures for training. This provides the network with diverse vertebrae poses during training, increasing the robustness of the shape completion model. When adjusting the spine's curvature, it is vital to consider the spine's physical constraints. \revisionRemove{as previously published~\cite{azampour2023anatomy}} \revisionAdd{A step-by-step algorithm of how we achieve multiple spine curvatures through realistic spine model deformations can be found in the supplementary material. This algorithm follows the approach proposed by Azampour et al.~\cite{azampour2023anatomy}}


\subsubsection{Generation of \gls{US}-consistent partial views of the spine}\label{subsection:Spine Surface Generation}
\begin{figure}[t]
    \centering
    \includegraphics[width=0.8\textwidth, keepaspectratio]{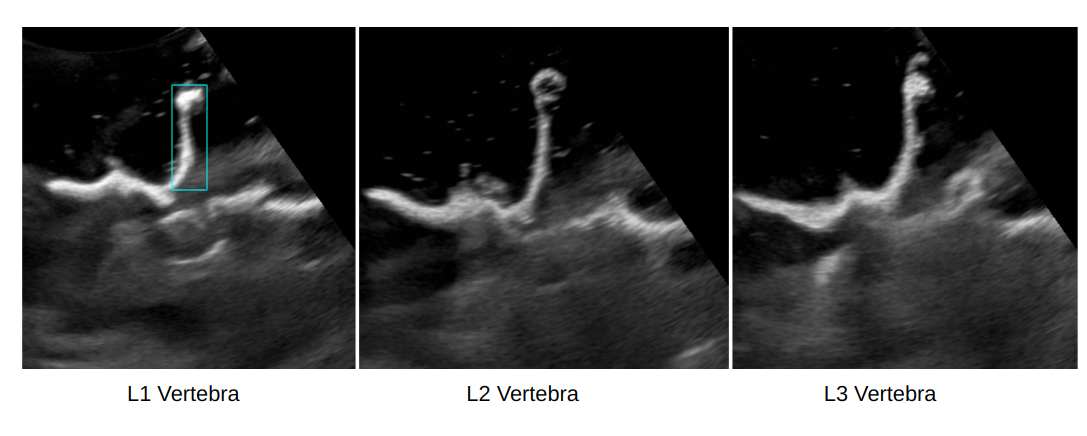}
    \caption{US scan of L1, L2 and L3 vertebrae levels of a spine phantom. These images exemplify the partial view of the vertebral arch as well as US-specific artifacts. We can see the effects of acoustic shadowing in the partially visible spinous process, highlighted with a bounding box on the left image.}\label{fig:US_scan_phantom}
\end{figure}

Spine US scans, whether transverse or paramedian, typically display only the vertebral arch's surface. Figure~\ref{fig:US_scan_phantom} showcases partial vertebrae in US, displaying verterbrae L1, L2, and L3 of a lumbar phantom. Structures like the spinous process, the laminae, the articular processes, and the transverse processes are only partially visible. Parts of these structures are rendered invisible due to a large angle of incidence between the direction of the US wave and the respective tissue. In other cases, they are occluded by the surrounding structures due to the effect of acoustic shadowing. Notably, the vertebral body is frequently fully shadowed.

Beyond acoustic shadowing, US exhibit scattering, causing minor displacements in the way some tissue appear on the image. This artifact amplifies noise and occlusions in an US vertebral view. In what follows, we will showcase how our technique produces partial spine views consistent with these US-specific characteristics.

\paragraph{Angle of incidence-aware ray-casting}

\gls{US} visibility hinges on the interaction and reflection of ultrasound waves with internal body structures. A pivotal factor is the angle of incidence — the angle at which the US wave hits the tissue. When this angle is below 90°, the US beam reflects, capturing and displaying the signal. Yet, at angles over 90°, especially when tissue boundaries align with the beam, the signal goes undetected, omitting the tissue interface from the display. For authentic, US-consistent \gls{PC}s, accounting for this phenomenon is vital.

We simulate the transversal US acquisition on spine meshes to produce partial views. Addressing the angle of incidence, we employ a technique that is cognizant of it. We strategically position the virtual rendering camera over each spinous process, casting rays to identify visible points. In this process, we compute the angle between each ray and the tissue plane and omit points with incidence angles of $\ge 90^{\circ}$. The impact of this technique is more significant degrees of occlusion, thereby enhancing the resemblance of the resulting \gls{PC} to a US view.

\paragraph{Account for Ultrasound Scattering}

To emulate US scattering — an effect where US beams register off-plane echoes —, we simulate off-plane signals by subtly shifting the spine perpendicularly to the incident ray direction and ray-casting it alongside the originally positioned spine. From this mesh, we then retain points unobstructed by the shift. The resultant \gls{PC}, exhibits more shadows, thus mirroring an ultrasound view of the spine.

\subsubsection{Masking spine into separate vertebrae views}~\label{subsection:masking_spine}
Our data generation pipeline concludes with dividing the spine into individual vertebra views, resulting in five vertebral \gls{PC}s serving as partial network inputs.
Segmenting spinal ultrasound into distinct vertebrae levels is challenging and prone to errors. To the best of our knowledge, no method can accurately differentiate between vertebrae levels in ultrasound images. Hence, our approach aims for realistic completions without relying on this specific information. To increase our method's robustness, we augment our data by performing neighboring cloud fusion. This process merges the \gls{PC}s from one vertebra with points from directly adjacent vertebrae.


\subsection{\revisionAdd{Vertebrae} Shape Completion}\label{subsection:network_architecure}
For the completion of 
\revisionAdd{vertebrae} shapes, we employ a probabilistic approach based on \gls{VRCNet}~\cite{pan2021variational}.
The shape completion pipeline consists of two networks that follow the variational autoencoder architecture. They are trained end to end using a composite loss function that incorporates two distinct components: the \gls{KL} divergence loss and the \gls{CL} as reconstruction loss. The two networks are (1) \gls{PMNet} and (2) \gls{RENet} (for details see supplementary material).

\gls{PMNet} employs probabilistic modeling to yield initial coarse completions by decoding global features. During training, it grasps the prior distribution of vertebrae shapes, capturing essential details about shape, size, and symmetry. At inference, the model refines the shape using observed data and the posterior distribution, allowing for patient-tailored completions that respect both general anatomical priors and unique characteristics of the individual's spine.

\gls{RENet} operates on both the partial and coarse-complete \gls{PC}s produced by \gls{PMNet}. Using an encoder-decoder design enhanced with self-attention modules, it can aggregate point features across various scales. This is crucial for detailed vertebral completions, preserving input nuances from the partial cloud while recovering specific occluded anatomical features of individual anatomies.

The proposed method generates the completed shape as a \gls{PC}. To better visualize the results, we apply a step of post-processing and generate a vertebral mesh based on Poisson Surface Reconstruction.  

\subsection{Datasets description }\label{datasets}

\textbf{Large Scale Vertebrae Segmentation Challenge 2020 Dataset:}
One dataset utilized in our study, referred to as the VerSe20, comprises abdominal CT scans that contain detailed annotations and classifications of vertebrae. Specifically, VerSe20 includes 125 lumbar vertebrae, evenly distributed across different levels, with 25 vertebrae per level. For our work, VerSe20 serves as the foundational dataset of our synthetic data generation pipeline.



\noindent\textbf{Paired US/CT patient data:}\label{datasets_patient}
The patient data comprises a total of two paired \gls{US}/\gls{CT} scans \cite{nagpal2015multi}. \revisionAdd{The ultrasound sweeps were obtained while the patient was in a sitting position, which is the standard pose for epidural injections. Through this data we assess the applicability of our method for this procedure.} 
To input this data into the shape completion network, we first perform a manual annotation of the bone in ultrasound, followed by a rough separation of the vertebrae. To generate the ground truth (GT) complete vertebral shape, we apply the automatic spine segmentation method proposed by Payer et al.~\cite{payer2020coarse}, and obtain vertebra-wise segmentations. 

\noindent\textbf{Phantom Dataset:}
To evaluate the shape and pose preservation of landmarks visible in the initial US, we use a lumbar spine phantom. This phantom consists of all five lumbar vertebrae as well as the intervertebral disks and the sacrum.   

\FloatBarrier

\subsection{Shape completion metrics}
\noindent\textbf{General Metrics:}
In our evaluation process, we utilized three key metrics: \gls{CD}, \gls{EMD}, and \gls{F1}. The \gls{CD}, widely employed in the computer vision community, calculates the point-to-point distance between two \gls{PC}s: one representing the completed shape and the other the ground truth shape. To enhance interpretability, we scaled our \gls{CD} values by a factor of $10^{4}$ following the approach of \gls{VRCNet}. \gls{EMD} measures the dissimilarity between two shapes by quantifying the minimum amount of work required to transform one shape into the other. Lastly, to address the impact of outliers, we incorporated an adapted version of the F1-score, as proposed by Knapitsch et al. \cite{knapitsch2017tanks}. This metric represents the harmonic mean of precision and recall, and serves as an additional measure of our methodology's performance. 

\noindent\textbf{Anatomy-specific Metrics:}\label{sec:anatomy_metrics}
Moving away from the general metrics, we introduce two anatomy-specific metrics.  
The spinous process is typically visible in ultrasound scans, making it a key reference point for our shape completion network. We aim to maintain its integrity and make sure it is placed at the appropriate location. To assess this, we calculate the Spinous Process Chamfer Distance (SP-CD) metric. This metric involves comparing two point sets generated by manually annotating the centerline of the spinous process surface in both the input and the completion. This measurement allows us to evaluate the fidelity of the spinous process preservation and placement in the completed shape.

Another anatomic landmark is the facet joint, which connects neighbouring vertebrae. 
To ensure that the facet joints are preserved in the 3D completion at the correct position, we measure the distance between the facet joint's center in the reconstruction and its correct location from the CT-based ground truth. 
\section{Experiments}
Our study begins with an evaluation of our proposed method 
\revisionAdd{and comparison of two shape completion approaches: the network described in \ref{subsection:network_architecure} and the approach proposed by the PCN work~\cite{yuan2018pcn}. This exemplifies the capability of our pipeline to integrate any, and therefore the most suitable point cloud based shape completion approach for the task at hand.} Next, we conduct an experiment dedicated to verifying how well the visible anatomical landmarks in \gls{US} are preserved in the completion. Lastly, we conduct two ablation studies, which aim to investigate the impact of the two US-related steps in the data generation pipeline, i.e. the \gls{US} physics and the neighboring cloud fusion, on the accuracy of our results. To evaluate the suitability of each model for shape completion in patient ultrasound images, we also evaluate using the paired \gls{US}/\gls{CT} patient dataset, the details of which are outlined in Section~\ref{datasets_patient}. Our analysis includes both quantitative and qualitative results for a comprehensive understanding of the outcomes.

%

\subsection{Experimental Setup of Our Method}
We split the VerSe20 dataset subjects based into 60\%-40\%-20\% for training, validation, and testing. For each experiment, we train for 100 epochs. The optimization uses the Adam optimizer with a learning rate set at 0.0001. For training, a batch size of 8 is utilized, whereas during testing, a batch size of 2 is employed. The training procedures are executed on an NVIDIA GeForce RTX 2080 GPU. The training durations for the proposed methodology and the two ablation studies are roughly 15 hours, 5 hours, 15 hours, respectively. During the inference stage, the shape completion process for a batch comprising two vertebrae, on average, takes 0.22 seconds.

\subsection{Anatomical Landmarks Preservation }
In evaluating our method, we place special emphasis on the vertebral arch, given its partial visibility in US. We want to ensure that the shape and pose of the anatomical landmarks in the vertebral arch such as the spinous process and the lateral process are preserved. To evaluate if these landmarks are preserved, we compute the anatomy-specific metrics (Sec.~\ref{sec:anatomy_metrics}) on our phantom dataset. We choose the phantom instead of the patient data for this experiment, since it facilitates the correct identification of the landmarks' pose. 

\subsection{Ablation Study}
In our ablation study, we systematically explore the impact of individual steps in synthetic data generation on accurate patient shape completion. \par

\noindent\textbf{Synthetic Data without considering US physics:} 
This experiment assesses the significance of incorporating \gls{US} physics~\cite{case1998ultrasound} into synthetic data generation
For this experiment, we do not consider \gls{US}-specific acquisition modalities nor \gls{US}-artefacts while generating the data. This translates to a simplified ray-casting process, omitting considerations of angle of incidence and completely bypassing the scattering step.\par
\noindent\textbf{Synthetic data without performing neighboring cloud fusion:} 
This experiment explores the network's performance on inaccurately separated vertebrae from patient data when trained solely on point clouds on which neighboring cloud fusion augmentation has not been performed. These point clouds therefore contain only points relevant to the specific vertebra without including points from neighboring structures. To achieve this, we omit the masking step in the data generation pipeline and, based on the \gls{CT} annotations generate vertebrae \gls{PC}s that are meticulously separated from neighboring vertebrae.

\section{Results}
\subsection{Evaluation of proposed methodology on synthetic and patient data}

\begin{figure}[t]
    \centering
    \includegraphics[width=\textwidth, keepaspectratio]{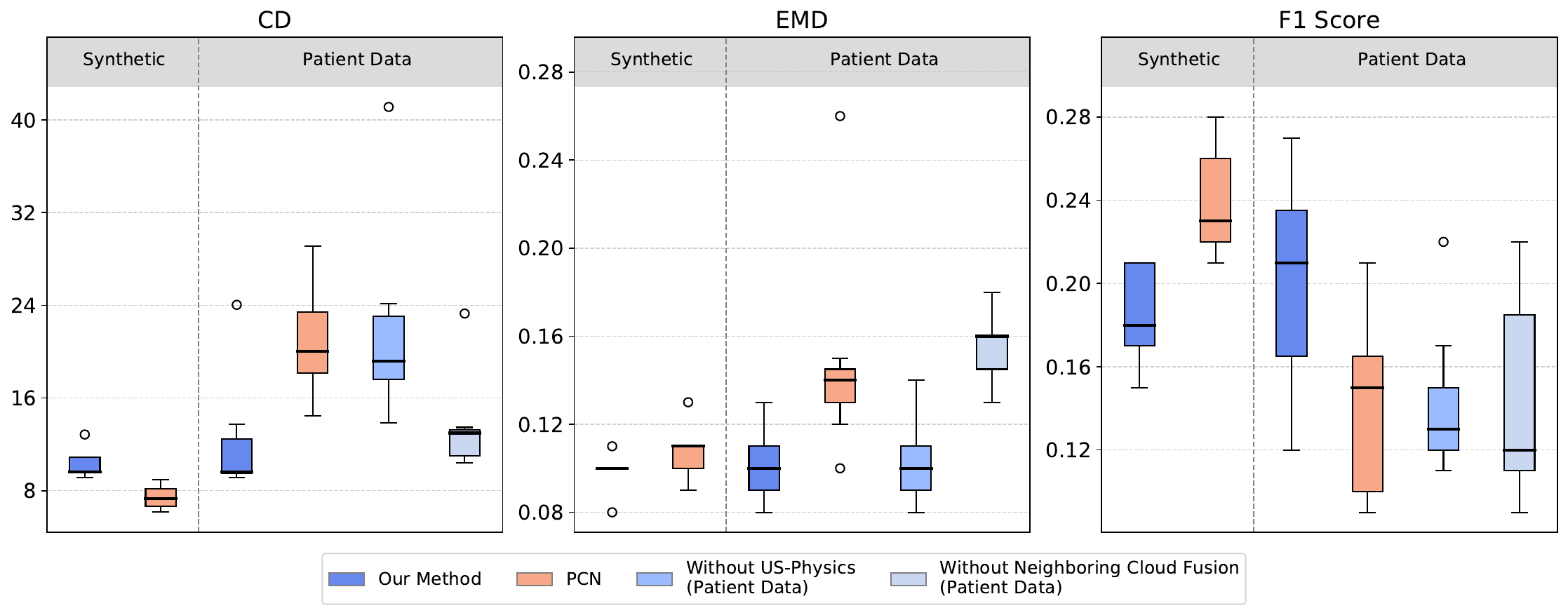}
    \caption{Performance comparison (in terms of Chamfer Distance (CD), Earth's Mover's Distance (EMD) and F1-Score) 
    \revisionAdd{of our full pipeline with two different shape completion approaches (VRCNet (blue) and PCN (orange))} on synthetic and patient data, as well as results of the ablation studies.}
    \label{fig:box_plot_results}
\end{figure}

The plots in Figure~\ref{fig:box_plot_results} compare the performance of the model trained on synthetic data both on the generated test set and patient data. Generally, the results on patient data show a larger variance and slightly lower accuracy. However, the differences of our method in all three metrics are relatively small, suggesting that our network can generalize from synthetic to patient data.

\revisionAdd{Additionally, we compare to PCN\cite{yuan2018pcn}, which, trained in the same manner, achieves comparable or even increased accuracy in the case of synthetic data. This demonstrates the interchangeability of the shape completion approach in our pipeline. However, unlike the proposed shape completion network, the PCN is not able to generalize well to the patient data. This, by comparison demonstrates the suitability of our chosen shape completion method for clinical applications.} 



\subsection{Preservation of landmark pose }
\noindent\textbf{Spinous Process:}
Table~\ref{landmarks_preserving} displays the dissimilarity between spinous process centerlines in the completion and in the input for each vertebral level. The accuracy only for this landmark is higher than the one for the complete shape. These results demonstrate the ability of the network to preserve the US-visible landmarks in the completion and reconstruct them at the correct position.  

\noindent\textbf{Facet Joints:}
The facet joint reconstruction accuracies measured as the distances between the center of the facet joint in the completion and GT are displayed in Table \ref{landmarks_preserving}. According to \cite{greher2004ultrasound}, an accuracy error of 5mm is still acceptable for a successful anaesthetic effect for facet joint injections. From our results, three out of four completed facet joint pairs 
would enable accurate injection delivery, while one pair (between L3 and L4) exceeds this threshold by at most 1.66mm.   

\begin{table}[t]
\caption{Facet joint reconstruction accuracy measured as the distance between the center of the facet joint in the completion and in the ground truth.}\label{landmarks_preserving}
\begin{tabularx}{\textwidth}{@{}YYYY@{}}
\toprule
\textbf{Vertebra Level }  & \textbf{ SP-CD} & \textbf{ Left Facet Joints Dist(mm)} & \textbf{ Right Facet Joint Dist(mm)}  \\ 
\midrule
L1   & 6.81   & 4.50  & 5.19\\
L2    & 2.00   & 2.64 & 4.87   \\ 
L3    & 2.88   & 4.97 & 3.46  \\ 
L4    & 6.09   & 6.45 & 7.66   \\ 
L5    & 5.88   & - & -  \\ 
\bottomrule
\end{tabularx}
\centering
\label{table:overview_experiments}
\end{table}

\subsection{Ablation Studies }

\begin{figure}[t]
    \centering
    \includegraphics[width=0.7\textwidth, keepaspectratio]{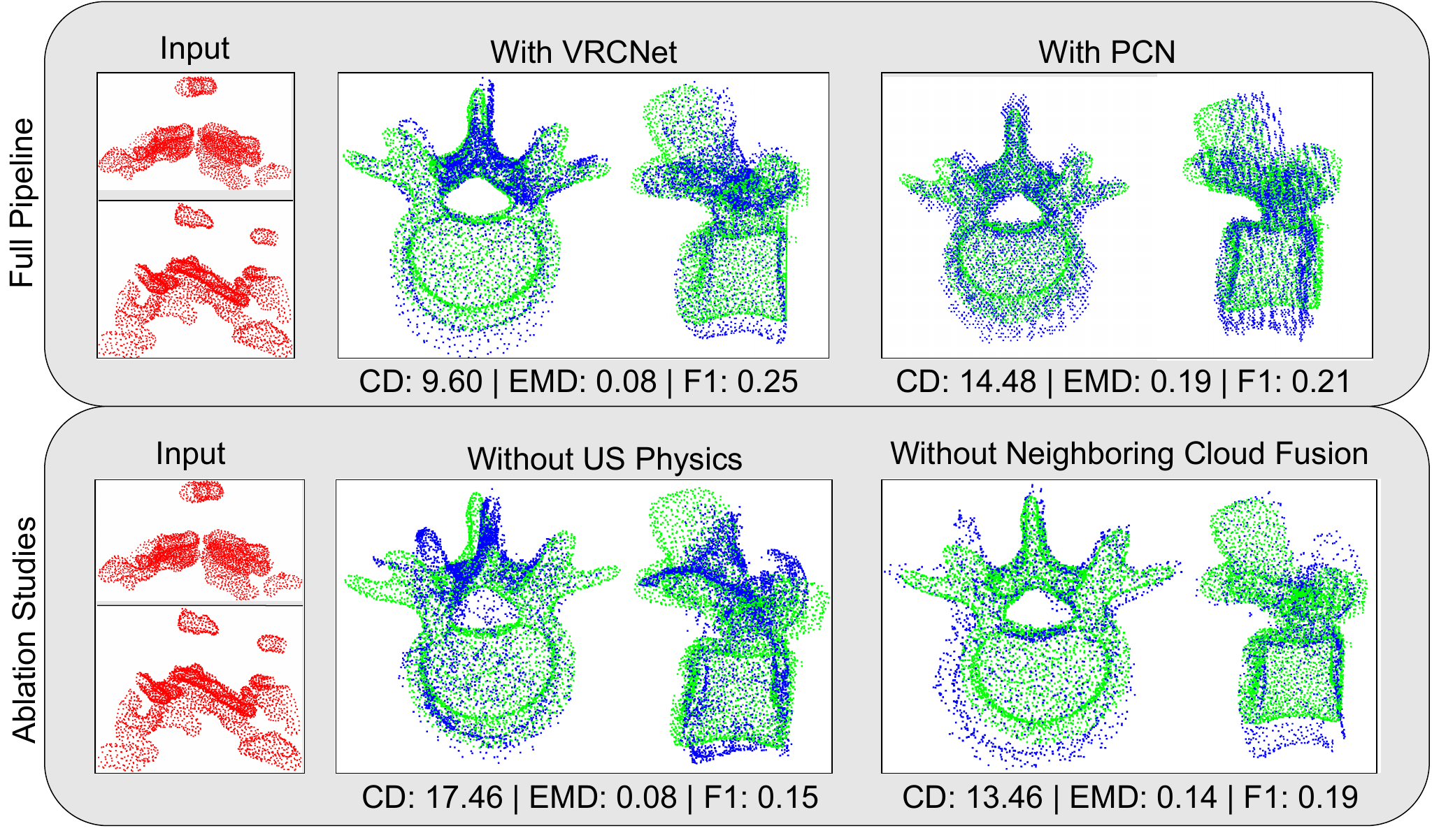}
    \caption{Patient data results obtained with \revisionRemove{our proposed method and} \revisionAdd{the full pipeline comparing two shape completion networks as well as two ablation studies.} \revisionRemove{two networks trained without considering US Physics and without performing neighboring cloud fusion.} Given our partial \gls{PC} input (red), We compare the reconstruction (blue) with the ground truth (green) and report three metrics: CD, EMD, and F1. We visualized the input and each completed shape PC from two views along the frontal and longitudinal axes.}
    \label{fig:qualitative_results}
\end{figure}

Figure~\ref{fig:box_plot_results} reports the quantitative results of the ablation studies on the patient dataset. Parallely, Figure~\ref{fig:qualitative_results} show examples of the qualitative results of our completions. 


Extensive qualitative results can be found in the supplementary material.  

\noindent\textbf{Results without US Physics:}
Considering the physics of US during the generation of synthetic data improves the accuracy of shape completion on patient data. As measured by the \gls{CD}, the accuracy of all completions increases. Specifically, we observe a median difference of 6.79 in the CD metric, indicating a noticeable improvement. This difference is also reflected in the other two scores, however, with a smaller median difference.

Qualitatively, omitting US physics simulation during training data generation leads to completion with unwanted points in the vertebral spinal canal, the area that houses the spinal cord. Additionally, important landmarks such as transverse processes and facets are missing in certain completions, for example, the transverse processes in Figure~\ref{fig:qualitative_results}. Furthermore, the completed shapes of the ablated model resemble less the GT shape, which can be particularly observed when looking at the vertebral body.  

\noindent\textbf{Results without neighboring cloud fusion:}
The proposed method outperforms the ablation model, an aspect which is reflected in all three metrics. The largest median difference of 0.06 is observed for the EMD score. 

In terms of qualitative assessments, the completions of the ablation model are relatively sparse. This is reflected in the low F1 values of this experiment. Moreover, the resulting completions contain multiple points in the spinal canal. 

\FloatBarrier
\section{Discussion and Conclusion}
In this work, we present a novel technique that addresses the challenge of completing anatomical structures given partial visibility in 3D US. Our method leverages synthetic data that considers US physics and artifacts, ensuring consistency with the partial display of anatomy in US. Moreover, generating this data considers process-specific augmentations such as curvature deformations and neighboring cloud fusion. We specifically apply our shape completion approach to the realm of US-based spine investigation. In this context, the proposed method completes the shapes of vertebrae. We demonstrate the generalizability of the proposed method to patient data, although trained only on synthetic data. This successful generalization emphasizes that our data generation process is realistic and US-consistent.

First of its kind, our proposed approach is capable of completing the shape of the vertebrae without prior patient-specific information, given only the US scan. This is particularly relevant in situations where a diagnostic CT scan is either unavailable or acquiring one is restricted due to factors like radiation concerns, for example, in the case of epidural injections.

The obtained results show promising outcomes, indicating the potential for further exploration in this area. Notably, enhancements in accuracy could be achieved by incorporating additional parameters such as vertebral level or patient BMI. These details could offer valuable contextual cues for the method, aiding in a more precise estimation of the vertebral shape. To advance towards highly accurate, patient-specific outcomes, the introduced approach could be refined during the testing phase by including the patient's CT scan data. This would provide precise information about the vertebra shapes, improving the performance of the US-based completion. 

\revisionAdd{
In clinical settings, highly accurate, patient specific completions would enable integrating the method into the workflow of spine injection surgeries. An ultrasound-based navigation system that displays the complete vertebral anatomy can assist surgeons in needle placement. For instance, it could help identify the level of the currently visualized vertebra in ultrasound, a very challenging and error-prone task. However, the final injection site confirmation would still rely on the original ultrasound. To optimize this guidance system, it's important to explore suitable rendering techniques and devise an adequate real-time component for use in the operating room. 
}

\revisionAdd{One current limitation of our work is the fact that it focuses on a single anatomy for shape completion. The ultrasound scan, which includes information about surrounding tissues, organs, and structures, is not fully utilized in the completion process. Incorporating this information could enhance the accuracy, providing cues about the size, pose or even abnormalities of the vertebral bodies, a structure not often captured in an ultrasound scan. 
This concept could be gradually extended to other regions of the spine, then to all types of rigid anatomies. Subsequently, devising appropriate methods to model and handle even deformable anatomies would be a relevant research direction.
}    

The proposed method relies on certain reference structures, \revisionAdd{such as the spinous process}, to be correctly segmented in US. This makes our method prone to errors if these structures are absent in the input \gls{PC} or wrongly segmented. Furthermore, the scope of our study was limited by the size of our dataset. While our research successfully demonstrated a proof of concept, a more comprehensive evaluation of the proposed method's capabilities necessitates a larger dataset, in particular paired US/CT patient data. A broader, large-scale study would provide a more thorough understanding of the method's performance across diverse scenarios, such as pathologies, different US acquisition protocols or quality, and further validate its effectiveness. 




In conclusion, the proposed method improves the interpretation of US images \revisionAdd{by enhancing the visualization of anatomic structures in US scans.} 
Mimicking how clinicians envision 3D anatomy, \revisionAdd{it incorporates} 
prior knowledge of the shape of the target structures, and \revisionAdd{considers} 
the physics of US imaging. \revisionAdd{In clinical practice, this technology could facilitate experts to rapidly and intuitively gain better understanding of the anatomy without the need for additional imaging modalities.} As an exemplary application, our method completes occluded vertebrae in US spine scans. We show that using synthetic 3D spinal views that consider the nature and artifacts of US imaging for training yields a model that provides consistent results on synthetic and clinical data. Notably, our approach maintains crucial anatomical landmarks in 3D completion, like the spinous process and the facet joints. Overall, this work shows a high potential for detailed lumbar vertebrae visualization and, ultimately, a path to explore towards the replacement of X-ray imaging for spine diagnosis and intervention.


\bibliography{sn-bibliography}

\newpage
\begin{appendices}

\section*{Supplementary Material}
This supplementary material (Online Resource 1) includes further details on implementation, datasets and a comprehensive overview of our qualitative results.

\section{Implementation Details}\label{secA1}

\subsection{Synthetic data generation pipeline}

\subsubsection{Angle of incidence-aware Raycasting}
\begin{figure}[h]
    \centering
    \includegraphics[width=11cm, keepaspectratio]{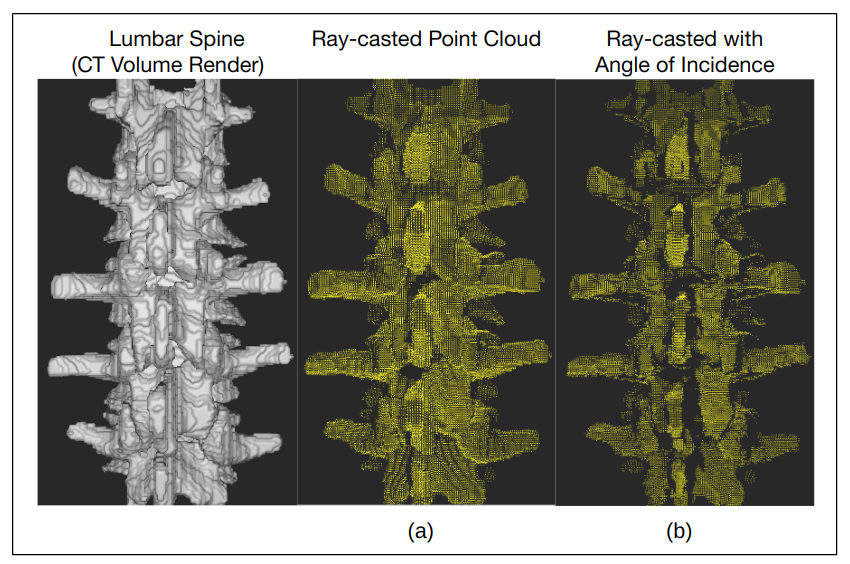}
    \caption{Comparison of spine mesh ray-casting when (a) the angle of incidence is not considered (b) the angle of incidence is considered. The resulting point cloud contains more shadows and is, therefore, more similar to the US view of the spine.}\label{fig:angle_of_incidence}
\end{figure}

To better understand the visual impact of the data generation pipeline on the resulting vertebral point cloud, we display a comparison of the spine mesh ray-casting with and without considering the angle of incidence in Figure \ref{fig:angle_of_incidence}. We observe that considering the angle of incidence leads to point cloud with more shadows, reflecting the shadowing effects in US.   

\FloatBarrier

\subsubsection{Account for Ultrasound Scattering}
\begin{figure}[h]
    \centering
    \includegraphics[width=0.8\textwidth, keepaspectratio]{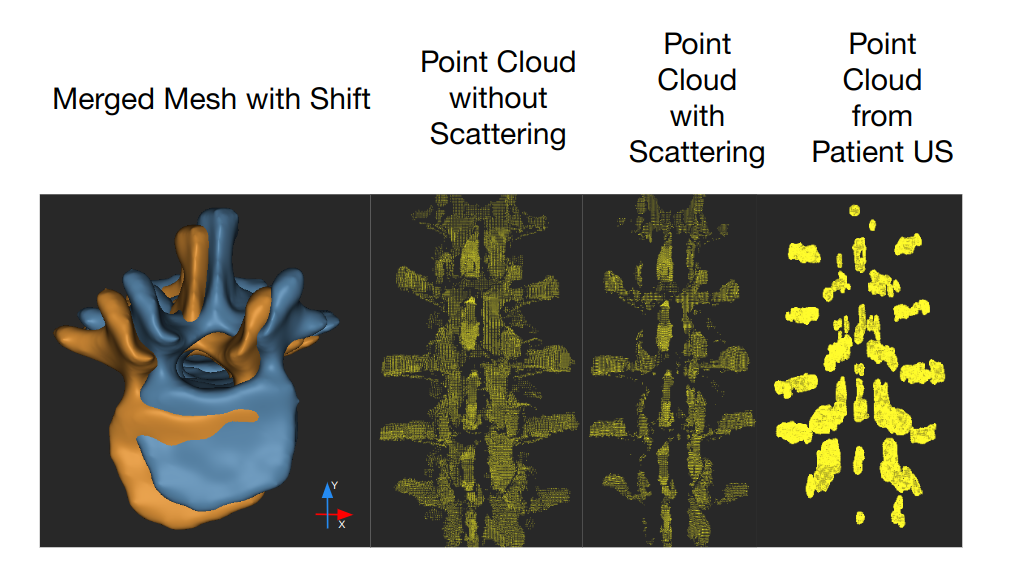}
    \caption{Left: Example of merge between centered in blue and shifted mesh in orange by -7mm along the lateral axis (x-axis) and -5mm along the anterior-posterior axis (y-axis). Right: Comparison of ray-casted point cloud with and without scattering. We observe that the scattered version displays more shadowed areas.   }
    \label{C3:fig:pcd_scattering}
\end{figure}

To simulate the effect of ultrasound scattering, we have empirically defined shift values symmetrically along the lateral axis and asymmetrically on the posterior-anterior axis, detailed in Table~\ref{C3:table:shifts_for_scattering}. The combined mesh of the centered (blue) and shifted spine (orange) is presented on the left side of Figure~\ref{C3:fig:pcd_scattering}. From this mesh, we retain points of the centered mesh unobstructed by the shift. The resultant point cloud, exhibiting more shadows, mirrors an ultrasound spine image, as visualized on Figure~\ref{C3:fig:pcd_scattering}'s right side. The occlusion extent and areas are shift-dependent. To diversify our synthetic output, we utilize all shift pairs from Table~\ref{C3:table:shifts_for_scattering}, yielding nine unique point clouds per dataset.

\begin{table}[h]
\begin{tabularx}{\textwidth}{@{}l|YYY@{}}
\hline
 &\textbf{Shift 1} & \textbf{Shift 2} & \textbf{Shift 3} \\ \hline
\textbf{Lateral axis (symmetrical)} & $\pm$5mm & $\pm$7mm & $\pm$10mm \\ \hline
\textbf{Anterior-posterior axis (asymmetrical)} & -1mm & -5mm & -10mm \\ \hline
\end{tabularx}
\centering
\caption{Shift values used to account for the effect of scattering in US. We use all possible pairs from these values, a total of 9 shifts per mesh for the data generation. }
\label{C3:table:shifts_for_scattering}
\end{table}

\FloatBarrier

\subsubsection{Masking spine into separate vertebrae views}
\begin{figure}[h]
    \centering
    \includegraphics[width=\textwidth, keepaspectratio]{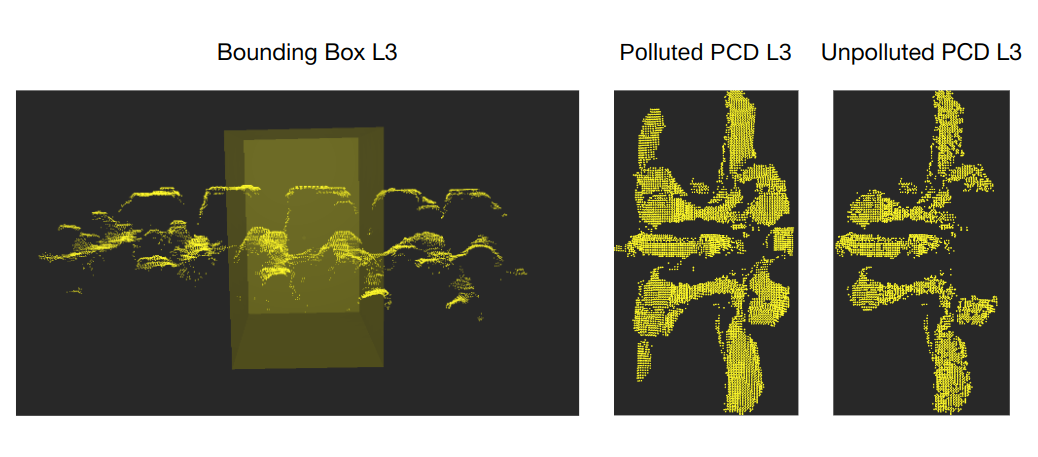}
    \caption{The spine point cloud is masked to obtain vertebra-wise point clouds. On the left, the masking process is visualized, in which we sequentially place a bounding box centered on each vertebra and select all points within. The middle and right image compare the results of this process which we call neighboring cloud fusion with the results of omitting it.}
    \label{C3:fig:vert_masking}
\end{figure}

To perform the neighboring cloud fusion augmentation, we sequentially place a bounding box centered on each vertebra's center of mass as visualized in Figure \ref{C3:fig:vert_masking}. We then extract all points within this bounding box to create the input point cloud. As exemplified in Figure \ref{C3:fig:vert_masking} where the box is centered on vertebra level L3, this technique allows us to collect and merge a small number of points from neighboring vertebrae (here L2 and L4) with the points from the current vertebra.   

\FloatBarrier
\subsection{Vertebrate Shape Completion Network Architecture}
\begin{figure}[h]
    \centering
    \includegraphics[width=\textwidth, keepaspectratio]{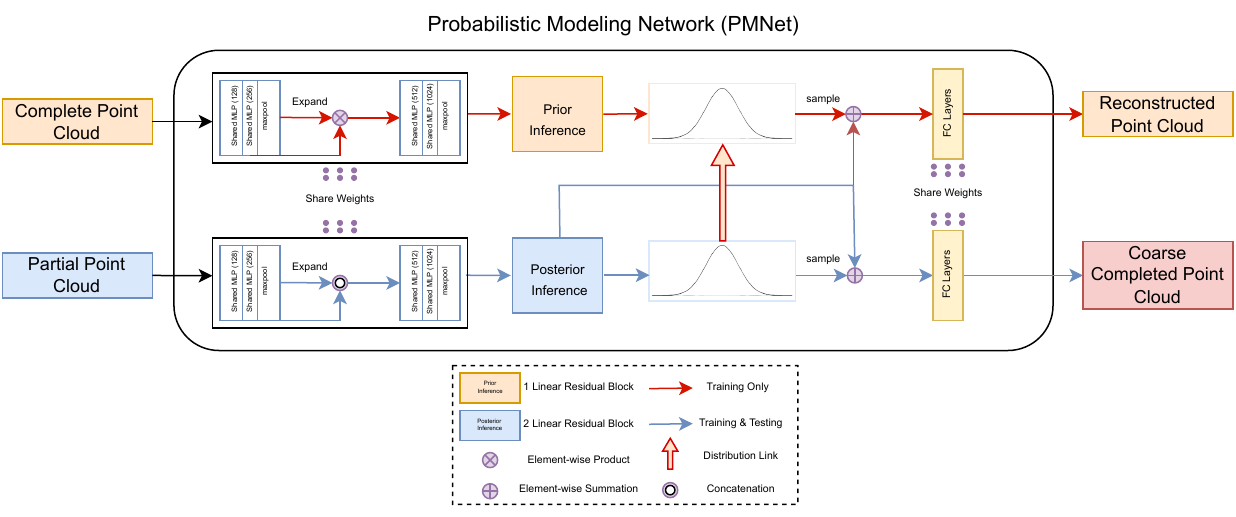}
    \caption{Overview of \gls{VRCNet}'s \gls{PMNet}, which generates the coarse complete point cloud at inference time based on the reconstruction latent distribution learned at training time. This figure was adapted from the original paper~\cite{pan2021variational}.}
    \label{C3:fig:VRCNET_PMNet}
\end{figure}

\begin{figure}[h]
    \centering
    \includegraphics[width=\textwidth, keepaspectratio]{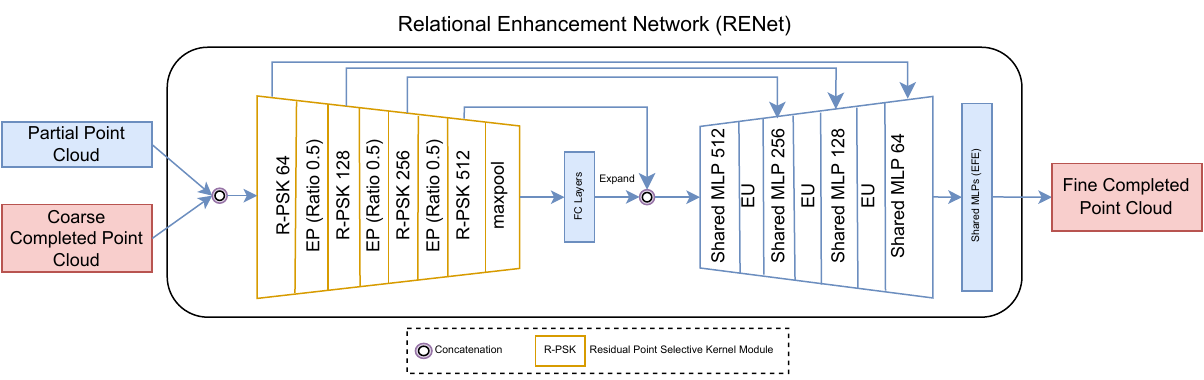}
    \caption{Overview of \gls{VRCNet}'s \gls{RENet}, network that generates the fine completion by using self-attention building blocks to recover local details. This figure was adapted from the original paper~\cite{pan2021variational}. }
    \label{C3:fig:VRCNET_RENet}
\end{figure}

The architecture of the shape completion method consists of two networks. (1) Probabilistic modeling network (PMNet) displayed in Figure\ref{C3:fig:VRCNET_PMNet} and (2) Relational Enhancement Network (RENet) shown in Figure\ref{C3:fig:VRCNET_RENet}.

(1) The PMNet consists of two paths. The first path, known as the reconstruction path, takes a complete shape as input and reconstructs it. Simultaneously, the completion path generates a coarse point cloud from the incomplete input. These pathways are designed in an autoencoder fashion. They share weights, and the reconstruction distribution is used to regularize the completion one. 
(2) The RENet is designed as an encoder-decoder structure with self-attention building blocks.

\FloatBarrier

\subsection{Baseline Network as Point Completion Network}
We consider the PCN\cite{yuan2018pcn} as the baseline architecture for shape completion and compare our method to it. We train the network with the chamfer distance loss for 400 epochs with a learning rate of 0.0001. 

\section{Dataset details}

\subsection{Phantom Data}
\begin{figure}[h]
    \centering
    \includegraphics[angle=90,origin=c,width=0.6\textwidth, keepaspectratio]{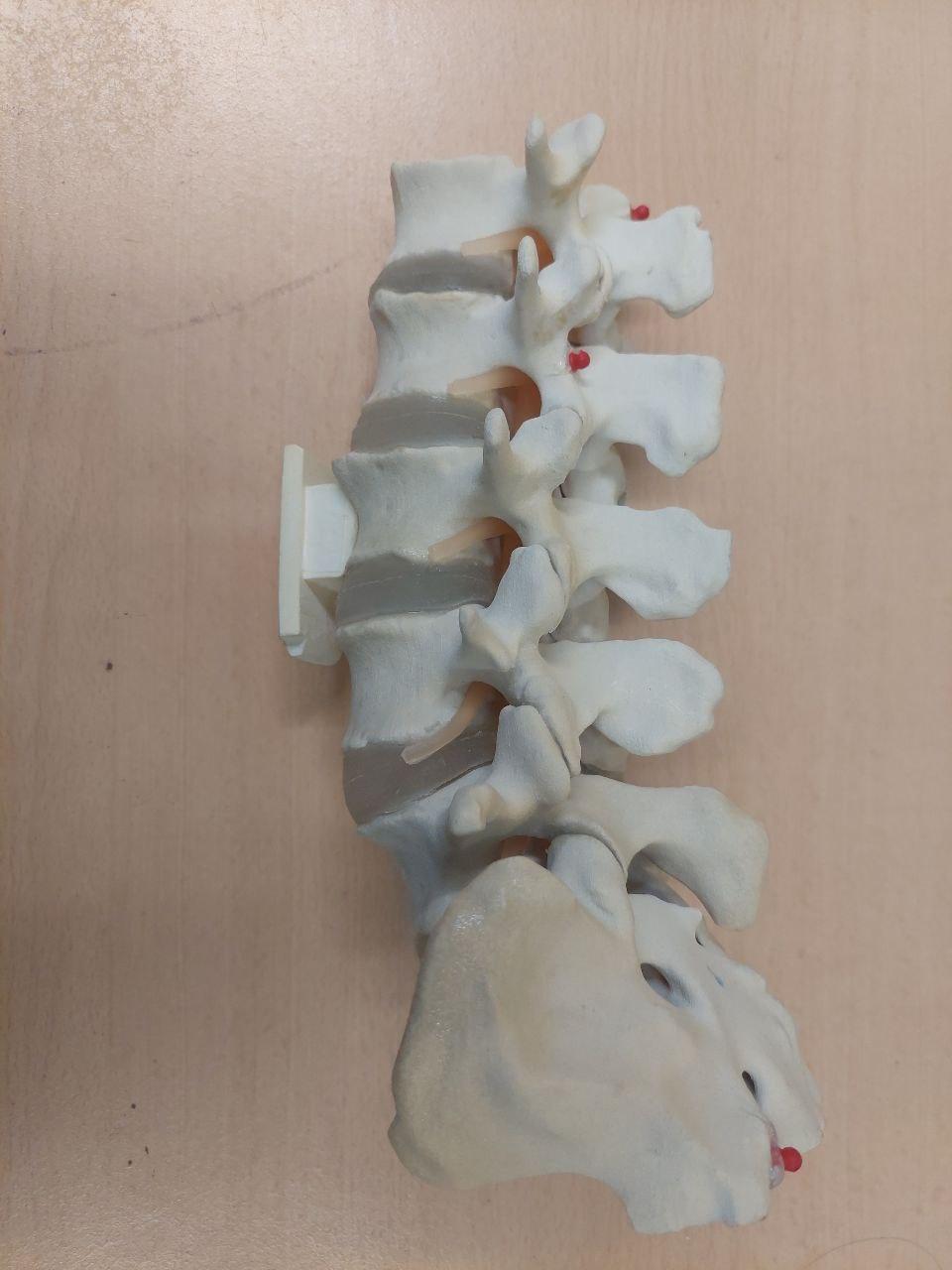}
    \caption{Anatomic model of lumbar spine used to demonstrate that the proposed method generates completions with correctly positioned vertebral landmarks. }
    \label{phantom_dataset}
\end{figure}

The lumbar spine phantom utilized in this work is shown in Figure~\ref{phantom_dataset}. It contains the five lumbar vertebrae as well as the sacrum and intervertebral disks. We acquired a 3D US scan with a transverse probe orientation by using ACUSON Juniper (Siemens Healthineers, Erlangen, Germany) with a 5C1 convex probe. This probe is mounted on a 7-axis robot of the model KUKA LBR iiwa 7 R800 manipulator (KUKA Roboter GmbH, Augsburg, Germany). 

\pagebreak
\section{Qualitative Results}

\begin{figure}[h]
    \centering
    \includegraphics[width=0.7\textwidth, keepaspectratio]{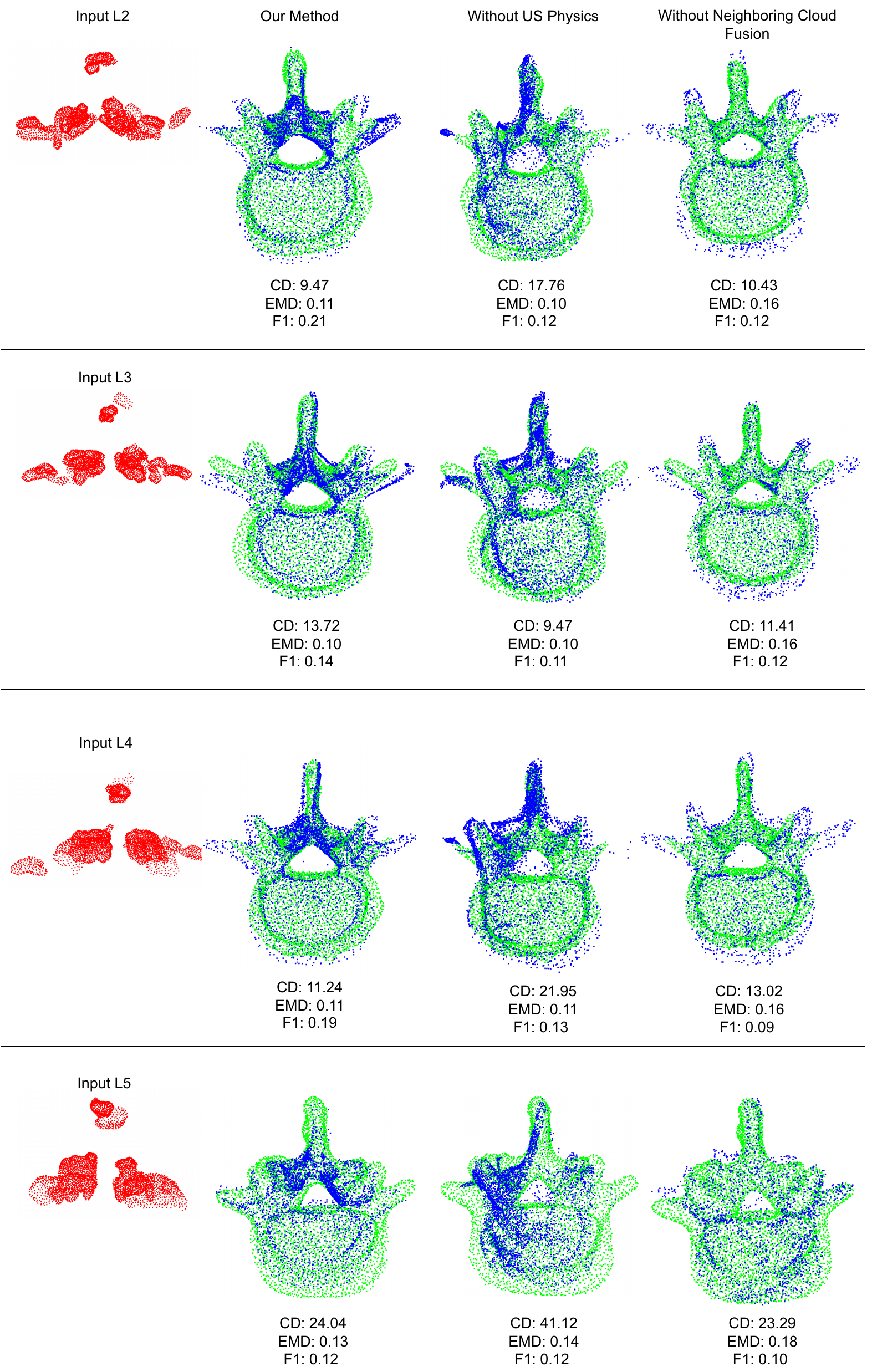}
    \caption{Qualitative results of the proposed method as well as ablation studies on Patient 1 \gls{US} data. Each row corresponds to one vertebral level. The first column displays the input to the network, while the others each show the completions in blue achieved by each network from Experiment 2 to Experiment 5 overlaid with the corresponding ground truth in green.  }
    \label{fig:qualitative_patient6}
\end{figure}

\begin{figure}[h]
    \centering
    \includegraphics[width=0.8\textwidth, keepaspectratio]{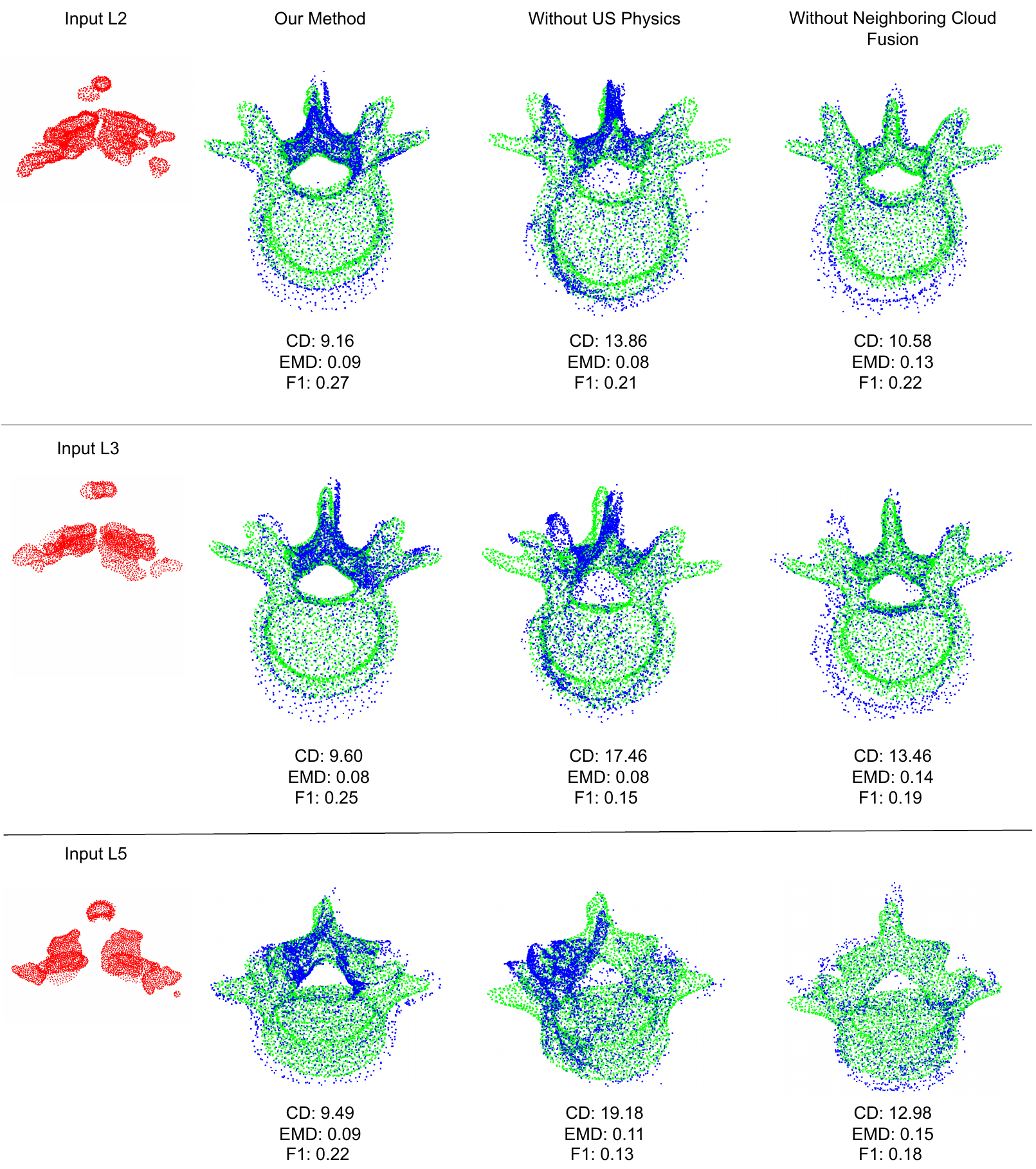}
    \caption{Qualitative results of the proposed method as well as ablation studies on Patient 2 \gls{US} data. Each row corresponds to one vertebral level. The first column displays the input to the network, while the others each show the completions in blue achieved by each network from Experiment 2 to Experiment 5 overlaid with the corresponding ground truth in green.}
    \label{C4:fig:qualitative_patient8}
\end{figure}

\begin{figure}[h]
    \centering
    \includegraphics[width=0.8\textwidth, keepaspectratio]{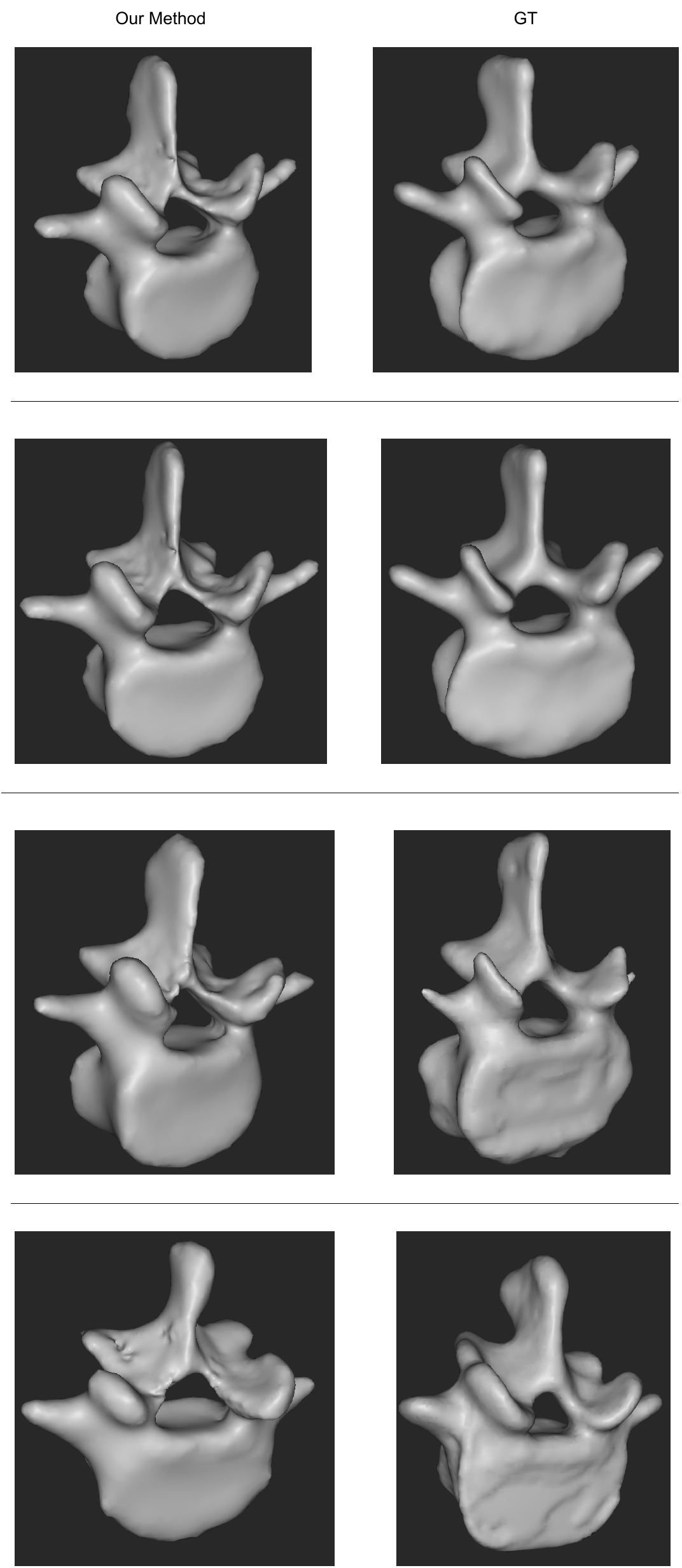}
    \caption{Visualization of post-processed results of the proposed method on Patient 1 \gls{US} data. Each row corresponds to one vertebral level, from L2 to L5. The first column displays the completed mesh while the second column shows the ground truth. }
    \label{fig:qualitative_patient6_meshes}
\end{figure}

\begin{figure}[h]
    \centering
    \includegraphics[width=0.8\textwidth, keepaspectratio]{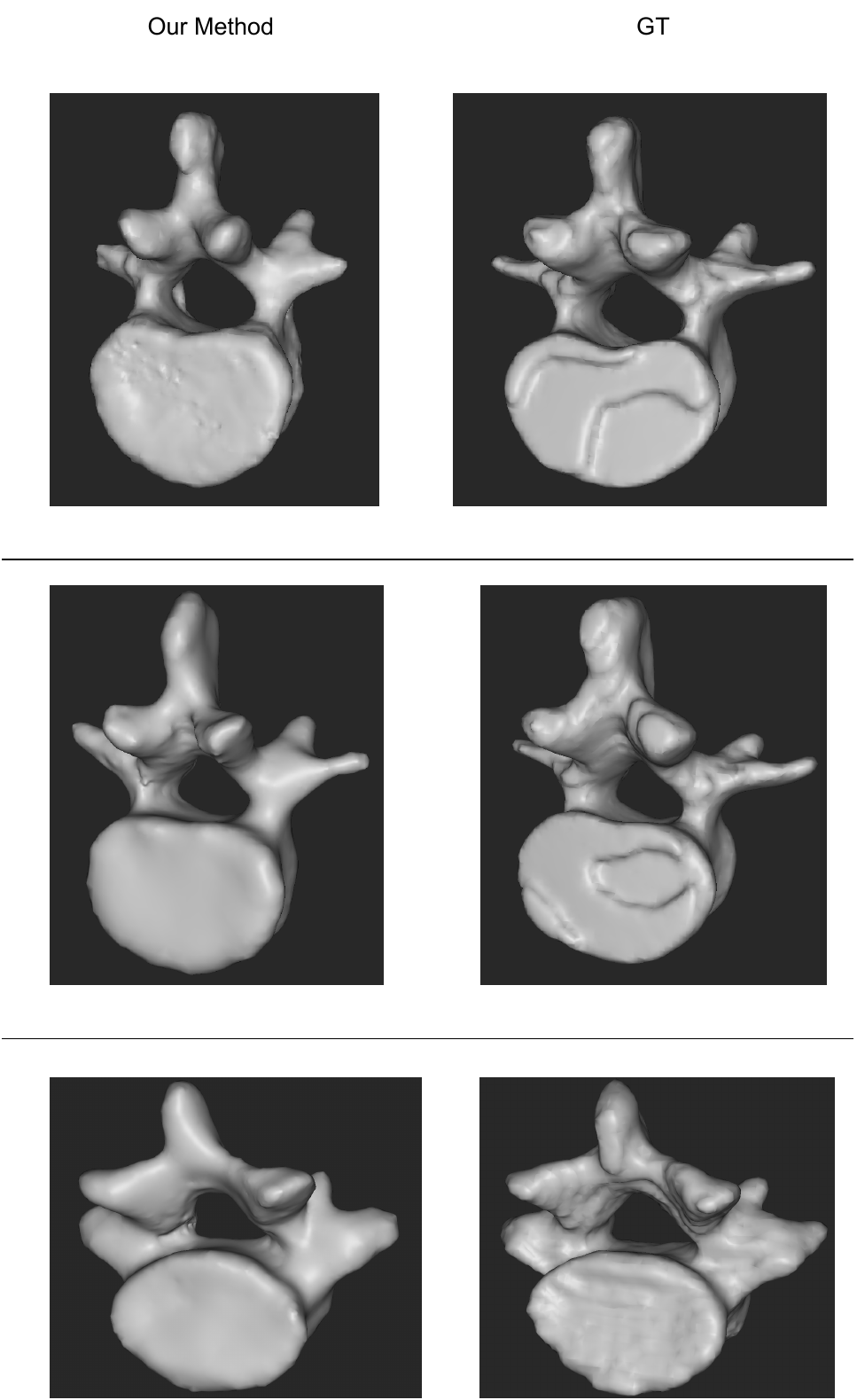}
    \caption{Visualization of post-processed results of the proposed method on Patient 2 \gls{US} data. Each row corresponds to one vertebral level,  L2, L3 and L5. The first column displays the completed mesh while the second column shows the ground truth. }
    \label{fig:qualitative_patient8_meshes}
\end{figure}

\begin{figure}[t]
    \centering
    \includegraphics[width=\textwidth, keepaspectratio]{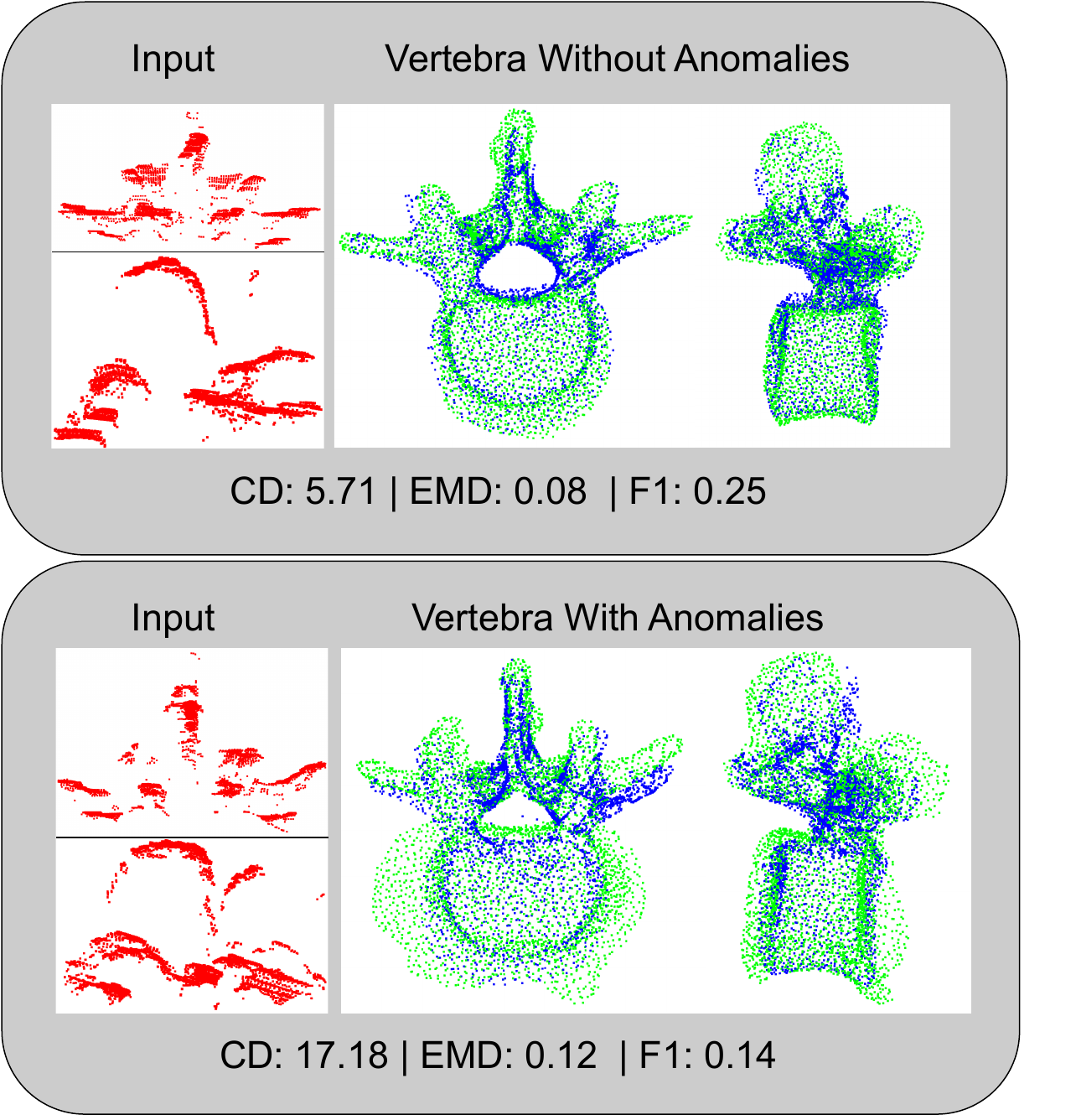}
    \caption{Shape Completion in the example of a vertebra with and without anomaly in the vertebral body. We notice that the anomaly occurs at the level of the vertebral body. Since no indication of this anomaly is present in the input to the shape completion network, the resulting completion reconstructs a normal vertebral body.  }
    \label{fig:pathologies_verse20}
\end{figure}



\FloatBarrier

\section{Data Generation Algorithm}

\begin{algorithm}
\caption{\revisionAdd{Synthetic Data Generation}}
\revisionAdd{
\begin{algorithmic}[1]
\State \textbf{Input:} Labelmap 
\State \textbf{Output:} Partial spinal pointcloud as seen in US
\State {\bf Extract mesh}
\Indent
    \State Run Marching Cubes algorithm on 3D label map
    \State Apply Gaussian Smoothing to reduce artifacts
\EndIndent
\State {\bf Deform mesh}
\Indent
    \State Use \Call{DeformSpine}{} algorithm for spine deformation \\
    \Comment{Refer to Spine Deformation Modeling algorithm}
\EndIndent
\State {\bf Ray-casting}
\Indent
    \State Position the virtual rendering camera over each spinous process
    \State Cast rays to identify visible points
    \State Compute the angle of incidence for each ray and the tissue plane
    \State Omit points with incidence angles of $\ge 90^{\circ}$    
\EndIndent
\State {\bf Simulate scattering}
\Indent
    \State Shift the spine mesh perpendicularly to the incident ray direction. The used shift values are found in Table~ \ref{C3:table:shifts_for_scattering}
    \State Perform ray-casting on both the original and shifted spine positions
    \State Retain unobstructed points from the shifted mesh    
\EndIndent
\end{algorithmic}
}
\end{algorithm}

\begin{algorithm}
\caption{Spine Deformation Modeling}  
\revisionAdd{
\begin{algorithmic}[1]
\State \textbf{Input:} Lumbar spine mesh
\State \textbf{Output:} Deformed spine model
\Procedure{DeformSpine}{}
    \State Define a physical model of the spine
    \Indent
        \State Model bones as rigid tissues in this model
        \State Model intervertebral fluids with springs
        \For{each vertebra $\mathcal{V}_i$ in $\{L_1, L_2, L_3, L_4, L_5\}$}
            \State Determine points on vertebral body and facets
            \State Compute centroid $c_i$
            \State Define inter-body and facet joint springs between the centroids
        \EndFor    
        \State \textbf{Spring parameters:}
        \Indent
            \State Inter-vertebral fluid stiffness: 500-1000 N/m
            \State Number of inter-vertebral springs: 400-800
            \State Facet joint stiffness: 8000 N/m
            \State Number of facet joint springs: 200-500
            \State Damping coefficients: 3 N/s (inter-body), 500 N/s (facet joints)
        \EndIndent
        \State Connect L1 and L5 to a still constraint
    \EndIndent
    \State Apply forces along anterior-posterior axis
    \State Generate deformations in multiple directions
    \State Apply varying forces to L1-L5 as per predefined intervals
\EndProcedure
\end{algorithmic}
}
\end{algorithm}
\end{appendices}

\end{document}